\documentclass[aps,prd,10pt,twocolumn,preprintnumbers,floatfix,nofootinbib]{revtex4-1}
\pdfoutput=1
\usepackage{graphicx}
\usepackage{bm}
\usepackage{times}
\usepackage{slashed}
\usepackage[usenames,dvipsnames,svgnames,table]{xcolor}
\usepackage[normalem]{ulem}
\usepackage{lipsum}
\usepackage{subfigure}
\usepackage{amsfonts}		
\usepackage{amsmath}		
\usepackage{amssymb}		
\newcommand{\be}{\begin{equation}}
\newcommand{\ee}{\end{equation}}
\newcommand{\bea}{\begin{eqnarray}}
\newcommand{\eea}{\end{eqnarray}}

\newcommand{\pbar}{\overline{p}}
\newcommand{\dbar}{\overline{D}}
\newcommand{\hebar}{\overline{^3\mathrm{He}}}

\begin{document}

\title{On the Origin of the Tentative AMS anti-Helium Events}

\author{Adam Coogan$^1$}
\email{acoogan@ucsc.edu}
\author{Stefano Profumo$^1$}
\email{profumo@ucsc.edu}

\affiliation{$^1$Department of Physics and Santa Cruz Institute for Particle Physics, University of California, Santa Cruz, CA 95064, USA}

\begin{abstract}
We demonstrate that the tentative detection of a few anti-helium events with the Alpha Magnetic Spectrometer (AMS) on board the International Space Station can in principle be ascribed to the annihilation or decay of Galactic dark matter, when accounting for uncertainties in the coalescence process leading to the formation of anti-nuclei. We show that the predicted antiproton rate, assuming the anti-helium events came from dark matter, is marginally consistent with AMS data, as is the antideuteron rate with current available constraints. We argue that a dark matter origin can be tested with better constraints on the coalescence process, better control of misidentified events, and with future antideuteron data. 
\end{abstract}

\maketitle

\section{Introduction}
Background-free processes are the Holy Grail of astrophysical searches for dark matter (DM): in numerous recent examples, ranging from the Galactic center excess to the positron fraction excess to the 3.5 keV line, possible DM signals have well-known, plausible astrophysical counterparts. Conclusively discriminating between a DM origin and a more prosaic astrophysical process is often challenging; however, the latter class of interpretations always carries the intellectual ``advantage'' of being preferred by Occam's razor.

A recent study, Ref.~\cite{Carlson:2014ssa}, argued that the discovery of even a single anti-helium-3 ($\hebar$) event at low-enough energies would be a virtually background-free signal of exotic physics; in that study, we also argued that it could be possible for DM annihilation or decay to produce $\hebar$ at  levels detectable by the Alpha Magnetic Spectrometer on board the International Space Station (AMS-02) \cite{ams} and by the future General Anti-Particle Spectrometer (GAPS) \cite{gaps} (see also Ref.~\cite{Cirelli:2014qia}).

Interestingly, late last year AMS publicly reported the tentative detection of a few proton-number $Z=-2$ events with a mass around the $\hebar$ mass, at a rate of roughly one event per year over the last five years, including a publicly-released event with a momentum of $40.3 \pm 2.9$ GeV \cite{AMStalk,Sokol240}. The AMS collaboration warns that with a signal-to-background ratio of roughly one event in $10^9$, very detailed instrumental understanding is paramount. Given the nature of the experiment, detector simulation studies are a key focus \cite{AMStalk}. In particular, the AMS Collaboration reports that it has dedicated so far 2.2 million CPU-days, produced around 35 billion simulated He events, and showed that ``the background is small'' \cite{AMStalk}. The Collaboration cautiously states that ``it will take a few more years of detector verification and to collect more data to ascertain the origin of these events'' \cite{AMStalk}. 

Event misidentification notwithstanding, in this study we consider the possibility that one or all of the tentatively detected antihelium events stem from DM annihilation or decay (for definiteness, we will hereafter focus on annihilation, but our results would apply directly to decaying DM models as well, {\em mutatis mutandis}) and assess the role of astrophysical backgrounds and nuclear physics uncertainties. The gist of our analysis is to (i) assume that the antihelium originates from DM and to (ii) calculate the resulting, unavoidable and, as we claim below, relatively robustly predictable antiproton and antideuteron associated flux, which we then (iii) compare to available data. We conclude that with current data, and given the uncertainty on the key parameter (the coalescence momentum) entering the formation of antinuclei in the hadronization of the annihilation products of dark matter particles, {\em the $\hebar$ events tentatively detected by AMS might indeed originate from Galactic dark matter}.

\section{Antihelium Production and Transport}

While cosmic-ray physics is notoriously ``messy'' because of uncertainties from Galactic cosmic-ray transport and solar modulation, flux ratios of cosmic-ray nuclei are largely free of such uncertainties and can be relatively robustly predicted. The key and by-far dominant uncertainty stems from the process underlying the formation of mass number $A>1$ (anti-)nuclei. The model which is customarily used relies on collider data from which a ``coalescence momentum'' $p_0^A$ is extrapolated. In the coalescence model, an antinucleus is assumed to form from constituent antinucleons if the antinucleons' 4-momenta lie within a sphere of diameter $p_0^A$. See Ref.~\cite{Carlson:2014ssa} for more details.

The coalescence momentum for (anti)deuterons is fairly well constrained by data from $e^+e^-\to\dbar$ from ALEPH at the $Z^0$ resonance \cite{ref3}, yielding for the coalescence momentum \cite{ibarra2013prospects} 
\begin{equation}\label{eq:deut}
    p_0^{A=2}=0.192\pm0.030\ {\rm GeV}.
\end{equation}

The $\overline{^3\mathrm{He}}$ coalescence momentum is very uncertain and not directly constrained by data. In Ref.~\cite{Carlson:2014ssa}, we used two different approaches to estimate $p_0^{A=3}$. In the first one, the scaling relation $p_0\sim\sqrt{B}$ \cite{ref8}, where $B$ is the nuclear binding energy, is used to produce
\begin{equation}
    p_0^{A=3}=\sqrt{B_{^3He}/B_{D}}p_0^{A=2}=0.357\pm0.059\ {\rm GeV}.
\end{equation}
In the second approach, Ref.~\cite{Carlson:2014ssa} used results from heavy-ion collisions at the Berkeley Bevalac collider which fit $D$, $^3$H and $^3$He coalescence momenta for several collision species (C+C up to Ar+Pb) with incident energies in the 0.4-2.1 GeV/$n$ range \cite{ref23, Baer:2005tw}. This results in the estimate
\begin{equation}
    p_0^{A=3}=1.28\ p_0^{A=2}=0.246\pm0.038\ {\rm GeV}.
\end{equation}
However, the coalescence momentum is known to depend significantly on the underlying scattering process since it must account for the whole poorly-understood process between hadronization and antinucleus formation. Since we have no data on $\hebar$ production in processes such as $p \pbar$ and $e^+ e^-$ which mimic DM annihilations better than heavy ion collisions, we use the binding energy scaling, taking $0.298\ {\rm GeV} \leq p_0^{A=3} \leq 0.416\ {\rm GeV}$ in what follows. While we expect our estimate based on $Z^0$ resonance data to be reliable for a DM particle with mass close to the $Z^0$ mass, further experimental data is required to understand whether it can adequately describe more massive annihilating DM particles \cite{pvdEmail}.

We fix the antihelium flux to a rate below that tentatively reproduces the AMS-02 events. Using the {\tt PYTHIA 8.156} \footnote{The extracted $p_0^A$ values and resulting antinucleus spectra depend on the choice of event generator \cite{PhysRevD.86.103536}. We estimate this choice introduces a factor of $2-4$ uncertainty on our final fluxes \cite{pvdEmail}.} Monte Carlo event generator we then reconstruct, for a given DM annihilation final state, the associated antideuteron and antiproton flux. The uncertainty on the antihelium coalescence momentum is propagated on the antideuteron and antiproton fluxes; the antideuteron fluxes additionally encompass the uncertainty in the deuteron coalescence momentum, as per Eq.~(\ref{eq:deut}). 


We model the antihelium propagation in the Milky Way by numerically solving the standard stationary, cylindrically symmetric, two-zone diffusion equation:
\begin{align}
    \frac{\partial f}{\partial t} &= 0 = \nabla\cdot(K(\vec{r}, T) \nabla f) - \frac{\partial}{\partial z} (V_c \operatorname{sign}(z)\ f) \notag\\
    &\hspace{0.5in} - 2 h \delta(z) \Gamma_{\rm int} f + Q_{\hebar}(T, \vec{r}).
\end{align}
In the equation above $f(\vec{r}, T)$ is the antihelium number density per unit kinetic energy and $\Gamma_{\rm int}$ is the interaction rate for antihelium with the interstellar medium (ISM).  

This diffusion equation is applied over a volume with radius fixed to 20 kpc and height $L$. The interstellar medium is contained in a disk with height $2h = 200\ {\rm pc}$ contained inside this volume. The diffusion coefficient is assumed to depend only on energy, and takes the form
\begin{equation}
    K(\vec{r}, T) = \frac{K_0 v}{c} \mathcal{R}^\delta,
\end{equation}
where $\mathcal{R} = p_{\rm GeV} / Z$ is the antihelium rigidity (using units of GeV for the $\hebar$ momentum) and $v$ is its velocity. Along with $K_0$, $\delta$ and $L$, the last parameter characterizing the model is the convection velocity $V_c$, which models the axially directed Galactic winds. The parameter values are fit using the boron to carbon ratio, giving the MIN/MED/MAX values listed in Table 1 of Ref.~\cite{ibarra2013prospects}. These models give $\pbar$, $\dbar$ and $\hebar$ fluxes that differ by a factor of $\lesssim 5$ at low energies, which is subdominant to the coalescence momentum uncertainties. More importantly for the present discussion, we calculate the {\em ratio} of $\pbar$, $\dbar$ to $\hebar$, which is largely insensitive to propagation parameters.

\subsection{Antihelium Interactions with the ISM}

The interaction rate between $\hebar$ and the ISM is given by
\begin{align}
    \Gamma_{\rm int} &= (n_{\rm H} + 4^{2/3} n_{\rm He}) v \sigma_{p,\hebar},
\end{align}
where $v$ is the $\hebar$'s velocity, $\sigma_{p, \hebar}$ is its interaction cross section with protons, and we assumed the helium and hydrogen gas cross sections are related by a geometric factor. We take the relevant densities in the Galactic Disk to be $n_{\rm H} = 1\ {\rm cm}^{-3}$ and $n_{\rm He} = 0.07 n_{\rm H}$. As discussed in detail in Ref. \cite{Carlson:2014ssa}, we bracket the uncertainty in $\sigma_{p,\hebar}$ using two methods. In MethodInel, we take $\sigma_{p,\hebar} = \sigma_{p,\hebar}^{\rm inel,non-ann}$, the inelastic, non-annihilating cross section for interactions with the ISM. With MethodAnn we instead use $\sigma_{p,\hebar} = \sigma_{p,\hebar}^{\rm tot} \equiv \sigma_{p,\hebar}^{\rm inel,non-ann} + \sigma_{p,\hebar}^{\rm ann}$, where the second term is the cross section for $\hebar$ to annihilate in collisions with protons.  While the two methods give very similar results above $\mathcal{O}(10 {\rm GeV})$, $\sigma_{p,\hebar}^{\rm tot} \gtrsim 2.5 \sigma_{p,\hebar}^{\rm inel,non-ann}$ at lower energies, leading to a flux about $40\%$ lower with MethodInn.

\begin{figure*}[!t]
    \center{\hspace*{-0.5cm}\includegraphics[width=0.95\textwidth]{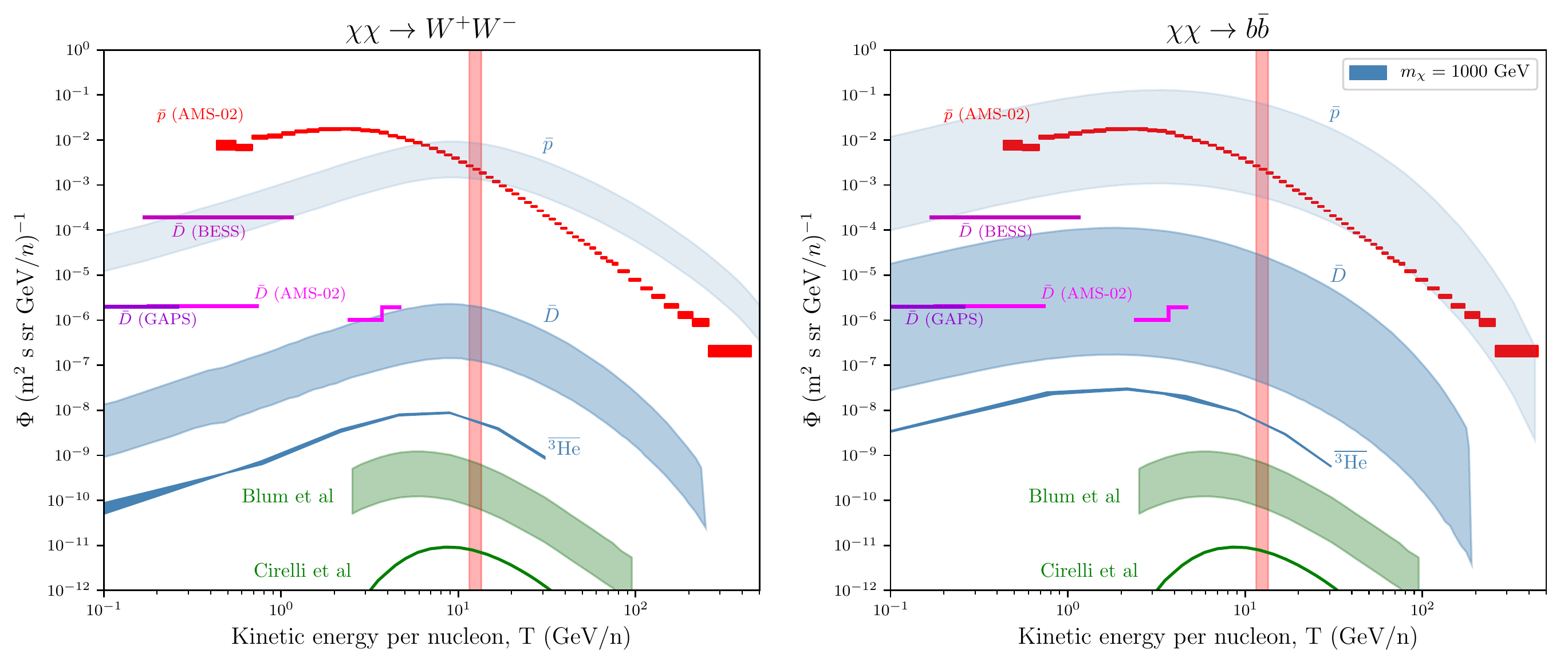}}
    \caption{The predicted antinucleon fluxes for a 1 TeV dark matter particle pair-annihilating into $W^+W^-$ (left panel) and $\bar b b$ (right panel), normalized to yield one $\hebar$ at 12 GeV/$n$ in five years. The spectra were computed using a Fisk potential of $\phi_F = 1.5$ GV, MethodAnn for $\hebar$-ISM interactions, and the MIN propagation parameters. The thickness of the lines reflects the uncertainties in the coalescence momentum for $\hebar$ formation in the case of antiprotons, and the combined uncertainty in the coalescence momentum relevant for $\hebar$ and $\dbar$ formation for the $\dbar$ flux. The green regions are the $\hebar$ background estimates from Blum et al's recent paper \cite{Blum:2017qnn} and from Cirelli et al \cite{Cirelli:2014qia}. The red rectangles indicate the AMS-02 antiproton flux data, while the pink and purple lines indicate the current (BESS \cite{PhysRevLett.95.081101}) and future (AMS-02 and GAPS \cite{Aramaki20166, Aramaki20161}) sensitivities to antideuterons. While the original GAPS satellite mission proposal projected a $\hebar$ sensitivity of $10^{-9}\ ({\rm m}^2 {\rm s\ sr\ GeV}/n)^{-1}$ for $0.1\ {\rm GeV}/n \lesssim T \lesssim 0.25\ {\rm GeV}/n$ \cite{gapsProposal}, the detector design has been changed and there is no planned satellite mission. There is no current $\hebar$ sensitivity estimate, though it is expected to be similar to the $\dbar$ sensitivity \cite{pvdEmail}.}
\label{fig:1_12}
\end{figure*}

\subsection{The Dark Matter Source Term}

The DM contribution to the antihelium flux is captured by the source term
\begin{align}
    Q_{\hebar}(\vec{r}, T) &= \frac{1}{2} \frac{\rho_{\rm DM}^2(\vec{r})}{m_\chi^2} \langle \sigma v \rangle \frac{dN_{\hebar}}{dT},
\end{align}
where $\rho_{\rm DM}$ is the DM density, $m_\chi$ is its mass, $\langle \sigma v \rangle$ is its thermally-averaged zero-temperature cross section and $dN/dT$ is the differential injection spectrum. $T$ indicates the kinetic energy per nucleon. Note that as in Ref.~\cite{Carlson:2014ssa} we neglect Coulombian barrier effects and obtain the total $\hebar$ yield by summing the direct $\hebar$ and $\bar{^3{\rm H}}$ ones; see the discussion in Ref.~\cite{Cirelli:2014qia} on this point). We assume a Navarro-Frenk-White DM density profile:
\begin{align}
    \rho_{\rm DM}(r) &= \left( \frac{r_s}{r} \right)^\alpha \frac{\rho_0}{(1 + r/r_s)^{\alpha+1}},
\end{align}
with inner slope $\alpha = 1$, scale radius $r_s = 24.42$ kpc and $\rho_0$ chosen such that $\rho(r_\odot) = \rho_\odot = 0.39\ {\rm GeV/cm}^3$. As discussed in Ref. \cite{ibarra2013prospects}, Einasto and cored-isothermal profiles give similar results. Since, as for transport, the uncertainty from halo profile choice on ratios of antinuclei is much smaller than the nuclear physics and propagation parameter uncertainties, we do not study them here.

\subsection{Solar Modulation}

After propagation through the Milky Way, heliospheric magnetic field alters the interstellar antihelium flux $\Phi^{\rm IS}_{\hebar}$. We account for solar modulation using the Force Field Approximation \cite{forcefield}, in which the flux for a nucleus with mass number $A$, proton number $Z$ and mass $m_A$ at the top of the atmosphere (TOA) is given by
\begin{align}
    \Phi_{A,Z}^{\rm TOA}(T_{\rm TOA}) &= \frac{2 A m_A T_{\rm TOA} + A^2 T^2_{\rm TOA}}{2 m_A A T_{\rm IS} + A^2 T^2_{\rm IS}} \Phi_{A,Z}^{\rm IS}(T_{\rm IS}),
\end{align}
where $T_{\rm TOA}$ ($T_{\rm IS}$) is the TOA (interstellar) kinetic energy per nucleon and $\Phi^{\rm TOA}_{A,Z}$ ($\Phi_{A,Z}^{\rm IS}$) is the TOA (interstellar) flux for the species. The TOA and interstellar kinetic energies per nucleon are related by $T_{\rm IS} = T_{\rm TOA} + e \phi_F |Z| / A$, where $\phi_F$ is the Fisk potential. We consider a range of values for $\phi_F$ from 500 MV to 1500 MV.

\subsection{From Fluxes to Counts}

Given a flux for a nucleus $\Phi_{A,Z}$ at Earth, the number of events at AMS is obtained from the corresponding acceptance $\mathcal{A}_{A,Z}(T)$ and exposure time $\mathcal{T}$ using
\begin{align}
    N &= \int_{T_{\rm min}}^{T_{\rm max}} dT\ \Phi_{A,Z}(T) \mathcal{A}_{A,Z}(T) \mathcal{T}(T).
\end{align}
No information about $\mathcal{A}_{\hebar}$ is publicly available, and we thus set it equal to the detector's geometric acceptance for $T > 0.5$ GeV/$n$, corresponding roughly to the lowest rigidity bin used in AMS's helium study. The exposure time is energy-dependent since the Earth's magnetic field shields low-rigidity particles. AMS-02's trajectory aboard the International Space Station passes through low latitudes where the geomagnetic cutoff is around $10$ GV (corresponding to an $\hebar$ particle with $T \approx 6$ GeV/$n$). We estimate this effect by taking $\mathcal{T} = (5\ {\rm years}) \times \varepsilon(T)$, where $\varepsilon(T)$ is the more optimistic of the two geomagnetic cutoff efficiency curves from Ref.~\cite{geoCutEff}.


\section{Results, Discussion and Conclusions}

\begin{figure*}[!t]
    \center{\hspace*{-0.5cm}\includegraphics[width=0.95\textwidth]{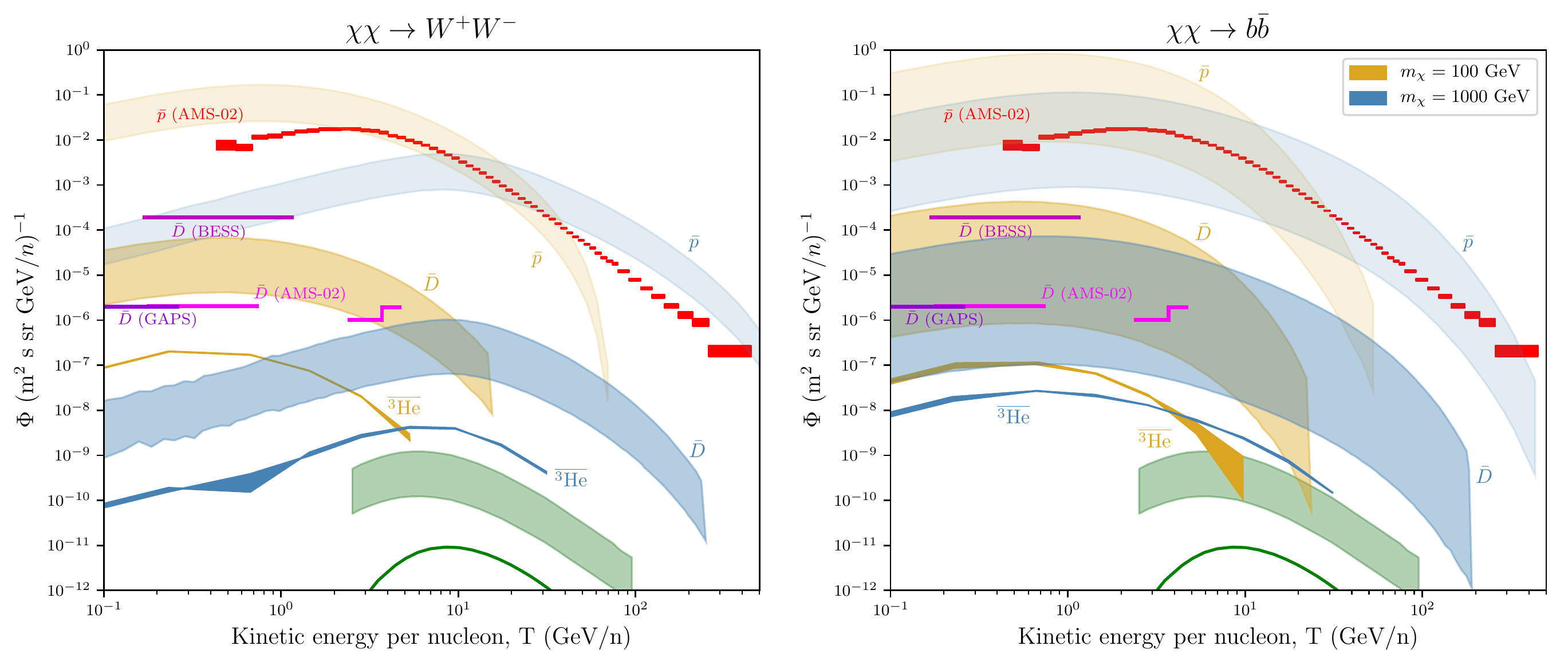}}
\caption{As in Fig.~\ref{fig:1_12}, but for the predicted antiproton and antideuteron fluxes for 100 GeV (yellow lines) and 1 TeV (blue lines) dark matter particles pair-annihilating into $W^+W^-$ (left panel) and $\bar b b$ (right panel), normalized to yield one $\hebar$ per year overall. Spectra are computed using $\phi_F = 500$ MV, MethodAnn and MAX propagation.}
\label{fig:1peryear}
\end{figure*}

In fig.~\ref{fig:1_12} we show the predictions for the antideuteron and antiproton fluxes for a 1 TeV dark matter particle self-annihilating into an unpolarized  $W^+W^-$ pair (left) and into a $\bar b b$ pair (right panel), generated using $\phi_F = 1.5$ GV, MethodAnn and the MIN propagation model, which is the most optimistic scenario for which AMS's $\pbar$ constraints are not violated. We normalize fluxes to obtain {\em one antihelium event} in five years with kinetic energy per nucleon in the $11.56\ \mathrm{GeV}/n \leq T \leq 13.5\ \mathrm{GeV}/n$ range, corresponding to the event publicly released by the AMS collaboration. The width of the predicted fluxes derives from the range of coalescence momenta for antihelium only (for the antiproton flux predictions) and for antihelium and antideuterium combined (for the antideuteron flux predictions). The red vertical band shows the kinetic energy range per nucleon of the putative event, and the two green bands at the bottom reproduce the $\hebar$ background estimates from Blum et al's recent paper \cite{Blum:2017qnn} and from Cirelli et al \cite{Cirelli:2014qia}. 

The figure illustrates that even generously accounting for uncertainties in the coalescence process, antiproton fluxes are too large for the $W^+W^-$ annihilation final state (left panel), especially at high energy. However, for the $\bar b b$ final state, it is possible to marginally be consistent with antiproton data, even though an excess should appear around 50 GeV/n in the antiproton data over the secondary background. While a dark matter origin  for the $\bar b b$ final state is possible for various combinations of the propagation setups, $\hebar$-ISM interaction method and $\phi_F$, all require setting $p_0^{A=3}$ to its maximum value. The scenario shown in the figure requires a thermally averaged pair-annihilation cross section of $\langle \sigma v \rangle = 3.58\times10^{-23}\ \mathrm{cm}^3 / \mathrm{s}$, which is, in addition, in tension with constraints from gamma-ray observations of local dwarf-spheroidal (dSph) galaxies with the Fermi Large Area Telescope (LAT) \cite{fermidsph}, although the latter present some systematic uncertainty. Moreover, AMS would expect, across the entire available energy range, to observe about three $\hebar$ per year rather than one.

Fig.~\ref{fig:1peryear} makes a different assumption about the tentative antihelium events, and it shows results for masses of 100 GeV (yellow) and 1 TeV (blue). Here we assume that the {\em overall} (i.e. the integrated) antihelium event rate for $T > 0.5$ GeV/$n$ is one event per year, as the AMS collaboration roughly indicated. Once again, the uncertainties in the antiproton and antideuteron fluxes are driven primarily by the uncertainties in the coalescence processes for $\hebar$ and $\dbar$ formation. The spectra in this figure were computed using $\phi_F = 500$ MV, MethodAnn and MAX propagation.

As before, a 100 GeV DM particle annihilating into $W^+ W^-$ is unable to explain the single $\hebar$ per year event rate without violating AMS's $\pbar$ bounds.  Annihilation into $\bar b b$ bodes better to suppress constraints from antiproton fluxes. For 1 TeV we find that one can get one antihelium event per year at AMS without violating antiproton constraints or antideuteron constraints with only slight tension with antiproton constraints for 100 GeV.  For this mass the required cross section is $\langle \sigma v\rangle = 7.30\times 10^{-26}\ {\rm cm}^3/{\rm s}$ while for $m_\chi = 1$ TeV the cross section is $\langle \sigma v\rangle = 4.78\times 10^{-25}\ {\rm cm}^3/{\rm s}$. These cross sections are a factor of $\sim 2$ above the 95\% C.L. Fermi-LAT dSph bound \cite{fermidsph}. Factoring in uncertainties in the dark matter density distribution both in the Milky Way (as relevant for the annihilation cross section necessary to produce the $\hebar$ flux) and in the sample of dwarf spheroidal galaxies utilized in the Fermi-LAT analysis, while there undeniably exist some tension, the cross sections we invoke cannot be firmly ruled out.

It is important to note that the relatively flat antihelium spectrum expected from the $b\bar{b}$ final state means that an event rate of one $\hebar$ per year over all energy bins is compatible with the single event whose energy has been publicly released, at least for sufficiently large dark matter mass. For example, the probability of observing one antihelium particle with momentum $p = 40.3 \pm 2.9$ GeV is $21\%$ for the 1 TeV case.  

Assuming a DM explanation for the tentative $\hebar$ events, the antiproton flux from dark matter would contribute significantly to the total antiproton flux at higher energies, perhaps compatibly with a possible weak excess of energetic antiprotons \cite{Cuoco:2016eej,Cui:2016ppb}. AMS-02 and GAPS would also be {\em likely} to detect a significant amount of antideuterons, but {\em non-detection is also possible within the full range of values for the coalescence momenta}. 

Since the known $\hebar$ event is at relatively large momentum, the level of the astrophysical background is a possible concern. App.~A of Ref.~\cite{Cirelli:2014qia} examines the $\hebar$ background using the coalescence model (green curve in Figs.~\ref{fig:1_12} and \ref{fig:1peryear}). In contrast to our prescription, they define the coalescence momentum $p_{\rm coal} = 167\ {\rm MeV}$ as a cutoff on the 3-momentum difference between constituent antinucleons' and use the same value for computing $\dbar$ and $\hebar$ production rates. A full study of how uncertainty about the coalescence momentum impacts the $\hebar$ background estimate is beyond the scope of this work. We also show the recent background estimate from Ref.~\cite{Blum:2017qnn} (green region in Figs.~\ref{fig:1_12} and \ref{fig:1peryear}), which attempts to account for the coalescence momentum's center of mass energy dependence using an analysis tool from heavy ion physics. While this background estimate is 1-2 orders of magnitude larger than the one from Ref.~\cite{Cirelli:2014qia}, it is still be about an order of magnitude less than required to give one event per year at AMS. A secondary cosmic-ray origin for the reported events is therefore  unlikely.

Admittedly, for the example masses and cross sections we consider, there exists some tension with gamma-ray observations \cite{fermidsph}, under relatively restrictive and somewhat aggressive assumptions on the uncertainties in the dark matter halo density profile of local dwarf galaxies and the Galactic center.  More generous assumptions, however, would relax Fermi-LAT bounds relative to the $\hebar$ flux levels needed to explain the tentative AMS events, thus easily allowing for the pair-annihilation rate considered here \cite{Ackermann:2013yva}. Constraints from positrons and neutrinos are much weaker than constraints from gamma rays and antiprotons for all the scenarios we have considered. 

We note that the results we presented here are generic for any source of high-energy antinucleons from hadronization of a high-energy parton. Our discussion therefore also encompasses the possibility that the antihelium events stem from, for example, primordial black hole (PBH) evaporation \cite{Kiraly:1981ci,Turner:1981ez}. Since there is little difference between the $\hebar / \pbar$ ratio from heavy quarks, light quarks and gluons, the only difference between our scenario and a PBH origin comes from the spatial distribution of the source term. As this difference is subdominant to uncertainties in the coalescence process, the PBH conclusions should mirror what we found for the $\bar b b$ case here.

Finally, a much more exotic possibility is that the detected antinuclei were produced in distant antigalaxies and propagated across cosmologically significant distances \cite{Streitmatter:1996pn}.  Observations of the extragalactic gamma-ray background strong constrain the existence of nearby antimatter domains (see e.g. \cite{Steigman:1976ev}), but our results cannot rule out this possibility.

In conclusion, we showed here how the few antihelium events reported by the AMS Collaboration can, in principle, be ascribed to exotic processes possibly involving the annihilation or decay of dark matter particles with masses in the TeV range and pair-annihilating into a quark-antiquark pair. For large-enough values of the coalescence momentum for antihelium formation, the resulting antiproton flux is marginally compatible with data, and the antideuteron flux is below current available constraints.

A conclusive answer to the question of the nature of the AMS antihelium events will require better instrumental understanding and assessment of misidentified He events in the detector, a task which in turns begs for extensive detector simulations. Whether or not the dark matter hypothesis is viable depends, additionally, on a firmer determination of the antihelium coalescence momentum, which dedicated collider data could help improve upon. Finally, future AMS results on the flux of cosmic-ray antideuterons, and especially results from the future GAPS instrument, could shed more light on the origin of the antihelium events and corroborate or in some cases rule out a dark matter interpretation.

\begin{acknowledgments}
    AC and SP are partly supported by the US Department of Energy, grant number DE-SC0010107. We are very thankful to Philip von Doetinchem for clarifying GAPS' capabilities and comments on the coalescence model, to Veronica Bindi for highlighting the importance of the geomagnetic cutoff, and to Francesco D'Eramo for feedback on the manuscript.
\end{acknowledgments}

\bibliographystyle{apsrev4-1}

\bibliography{main}

\end{document}